\documentclass[manuscript]{aastex}

\usepackage{CJK}
\usepackage{psfig}

\newcommand{\E}[1]{\times 10^{#1}}
      \newcommand{\ps}{\,{\rm s}$^{-1}$}
    \newcommand{\Msun}{M_{\odot}}
\newcommand{\cm}{\,{\rm cm}}    \newcommand{\km}{\,{\rm km}}

\newcommand{\ncol}{N(H$_2$)}
\newcommand{\xray}{X-ray}
\newcommand{\ROSAT}{{\sl ROSAT}} 
\newcommand{\Chandra}{{\sl Chandra}} 
\newcommand{\du}{d$_{1}$}  

\newcommand{\vlsr}{V$_{\rm LSR}$}       \newcommand{\tmb}{T$_{\rm mb}$}
\newcommand{\twCO}{$^{12}$CO}  \newcommand{\thCO}{$^{13}$CO}
\newcommand{\CeiO}{C$^{18}$O} 
\newcommand{\snr}{{G127.1$+$0.5}}

\shorttitle{Molecular filament associated with SNR G127.1$+$0.5}
\shortauthors{Zhou et al.}

\begin{document}
\begin{CJK*}{UTF8}{gbsn}

\title{Discovery of a pre-existing molecular filament associated with supernova remnant G127.1$+$0.5}

\author{Xin Zhou (周鑫)\altaffilmark{1,2}, Ji Yang\altaffilmark{1,2}, Min Fang\altaffilmark{1,2}, and Yang Su\altaffilmark{1,2}}
\affil{$^1$Purple Mountain Observatory, CAS, 2 West Beijing Road, Nanjing 210008, China; xinzhou@pmo.ac.cn \\
$^2$Key Laboratory of Radio Astronomy, Chinese Academy of Sciences, Nanjing 210008, China
}


\begin{abstract}
We performed millimeter observations in CO lines toward the supernova remnant (SNR)~\snr. We found a molecular filament at 4--13~\km\ps\ consisting of two distinct parts: a straight part coming out of the remnant region and a curved part in the remnant region.
The curved part is coincides well with the bright SNR shell detected in 1420~MHz radio continuum and mid-infrared observations in the northeastern region.
In addition, redshifted line wing broadening is found only in the curved part of the molecular filament, which indicates a physical interaction.
These provide strong evidences, for the first time, to confirm the association between an SNR and a pre-existing long molecular filament.
Multi-band observations in the northeastern remnant shell could be explained by the interaction between the remnant shock and the dense molecular filament.
RADEX radiative transfer modeling of the quiet and shocked components yield physical conditions consistent with the passage of a non-dissociative J-type shock.
We argue that the curved part of the filament is fully engulfed by the remnant's forward shock.
A spatial correlation between aggregated young stellar objects (YSOs) and the adjacent molecular filament close to the SNR is also found, which could be related to the progenitor's activity.
\end{abstract}


\section{Introduction}

Core-collapse supernovae are thought to be commonly located in the vicinity of molecular clouds (MCs), due to their short lifetime after being born in giant MCs.
Therefore, the subsequent supernova remnants (SNRs) may encounter these MCs in their evolution.
In practice, it is difficult to confirm such interactions, because of the presence of extreme spectral and spatial complexities in studying the widespread molecular gases in the galactic plane where most SNRs are found.
To date, there are only 34 SNRs \citep[$\sim$12\% of known SNRs in our Galaxy, in contrast to about half of them indicated by][]{ReynosoMangum2001} confirmed to be interacting with MCs, and most of them were found through the detection of 1720~MHz OH masers \citep[catalogued by][see references therein]{Jiang+2010}. 
The shock-excited 1720~MHz OH maser is a signpost for SNR--MC interaction, but it is only available in particular physical conditions, i.e., in a moderate non-dissociative C-type shock \citep{WardleYusef-Zadeh2002}.
Since MCs are in different evolutionary stages, with different physical conditions at different positions within it, there should be various kinds of interactions between SNRs and MCs. It is important to study different kinds of interactions to figure out the role of SNRs in their parent MCs' evolutions, e.g., triggering the next generation of star formations.

SNR~\snr\ has been identified as a shell-type SNR by \cite{Caswell1977} and \cite{Pauls1977}.
Together with the central compact sources, they have been studied in many wavelengths: radio \citep{Salter+1978, Pauls+1982, Fuerst+1984, Joncas+1989, LeahyTian2006, Sun+2007}, optical \citep{Xilouris+1993}, and \xray\ \citep{Kaplan+2004}.
Its radio feature, which has been studied very well, shows a fairly circular shell peaked in north--northeast with the polarization {\textbf B}-vectors following the shell \citep{Sun+2007}, and a spectral index of 0.43$\pm0.1$ \citep{LeahyTian2006}.
Diffuse optical emission in the SNR has been detected \citep{Xilouris+1993}, but no diffuse \xray\ emission was found either by \Chandra\ observation toward the central region \citep{Kaplan+2004} or by the \ROSAT\ all-sky survey \citep{LeahyTian2006}.
No instantly associated neutron star has been detected in the SNR \citep{Kaplan+2004}.
The distance to the SNR is not clear either; the distance is mostly based on the SNR's assumed association with the open cluster NGC~559, which coincides in position with the SNR in previous works, hence the basic evolutionary facts of the SNR are still uncertain.
Nevertheless, previous studies have revealed the possibility of an SNR--MC interaction.
If such an association is established, we could better understand not only this particular SNR but also a different kind of SNR--MC interaction, i.e., without the detection of a 1720~MHz OH maser.
In this work, we present our CO line observations toward the SNR~\snr.
We describe the observations and results in Sections~\ref{sec:obs} and \ref{sec:result}, respectively. In Section~\ref{sec:discuss}, we discuss the interpretation and derive the characteristic parameters of the SNR--MC interaction and SNR evolution. The conclusions are summarized in Section~\ref{sec:conclusion}.

\section{Observations}\label{sec:obs}
The observations of millimeter molecular emissions toward SNR~\snr\ is part of the Milky Way Scroll Painting--CO line survey project\footnote{http://www.radioast.nsdc.cn/yhhjindex.php}
operated by Purple Mountain Observatory (PMO),
which were made from 2011 November to 2012 February with the 13.7~m millimeter-wavelength telescope of the Qinghai station of PMO at Delingha.\footnote{
Status Report on the 13.7 m Millimeter-Wave Telescope for the 2011-2012 Observing Season is available at http://www.radioast.csdb.cn/zhuangtaibaogao.php}
A superconductor--insulator--superconductor (SIS) superconducting receiver with a nine-beam array was used as the front end \citep{Shan+2012}.
Three CO lines were simultaneously observed, \twCO\ (J=1--0) at upper sideband (USB) and two other lines, \thCO\ (J=1--0) and \CeiO\ (J=1--0), at the lower sideband (LSB).
Typical system temperatures were around 210 K for USB and around 130 K for LSB, and the variations among different beams are less than $15\%$. 
The total of pointing and tracking errors is about $5''$. The half-power beam width (HPBW) is $52''$.
The main-beam efficiencies during the observing epoch were $\sim$44\% for USB with the differences among the beams less than 16\%, and $\sim$48\% for LSB with the differences less than 6\%.
We mapped a $100'\times 100'$ area that contains the full extent of SNR~\snr\ via on-the-fly (OTF) observing mode, and the data is meshed with the grid spacing of $30''$.
A Fast Fourier Transform (FFT) spectrometer with a total bandwidth of 1000 MHz and 16,384 channels was used as the back end.
The corresponding spectral resolutions were 0.17~\km\ps\ for \twCO\ (J=1--0) and 0.16~\km\ps\ for both \thCO\ (J=1--0) and \CeiO\ (J=1--0).
The baseline subtraction was performed with a linear fit. The average rms noises of all final spectra are about 0.5~K for \twCO\ (J=1--0) and about 0.3~K for \thCO\ (J=1--0) and \CeiO\ (J=1--0).
All data were reduced using the GILDAS/CLASS package.\footnote{http://www.iram.fr/IRAMFR/GILDAS}

The 1420~MHz radio continuum emission data were obtained from the Canadian Galactic Plane Survey \citep[CGPS;][]{Taylor+2003}.
The infrared (IR) photometric data were extracted from the Two-Micron All Sky Survey \citep[2MASS;][]{2MASS2006} and the survey of the {\it Wide-field Infrared Survey Explorer} \citep[{\it WISE};][]{WISE2010}.
The 2MASS survey provides data in the near-IR $JHK_{\rm S}$ bands with a 10$\sigma$ point-source detection level better than 15.8, 15.1, and 14.3~mag, respectively.
{\it WISE} achieved 5$\sigma$ point source sensitivities better than 0.08, 0.11, 1, and 6 mJy at 3.4, 4.6, 12, and 22$\mu$m, respectively. The angular resolutions are 6$''$.1, 6$''$.4, 6$''$.5, and 12$''$.0\ in the four {\it WISE} bands, respectively.
We used 2MASS $HK_{\rm s}$ bands and the first three {\it WISE} bands for point source analysis, and selected the point source with photometric error less than 0.1 mag for 2MASS data and signal-to-noise ratio greater than 5 for {\it WISE} data.

\section{Results}\label{sec:result}

\begin{figure*}[ptbh!]
\centerline{{\hfil\hfil
\psfig{figure=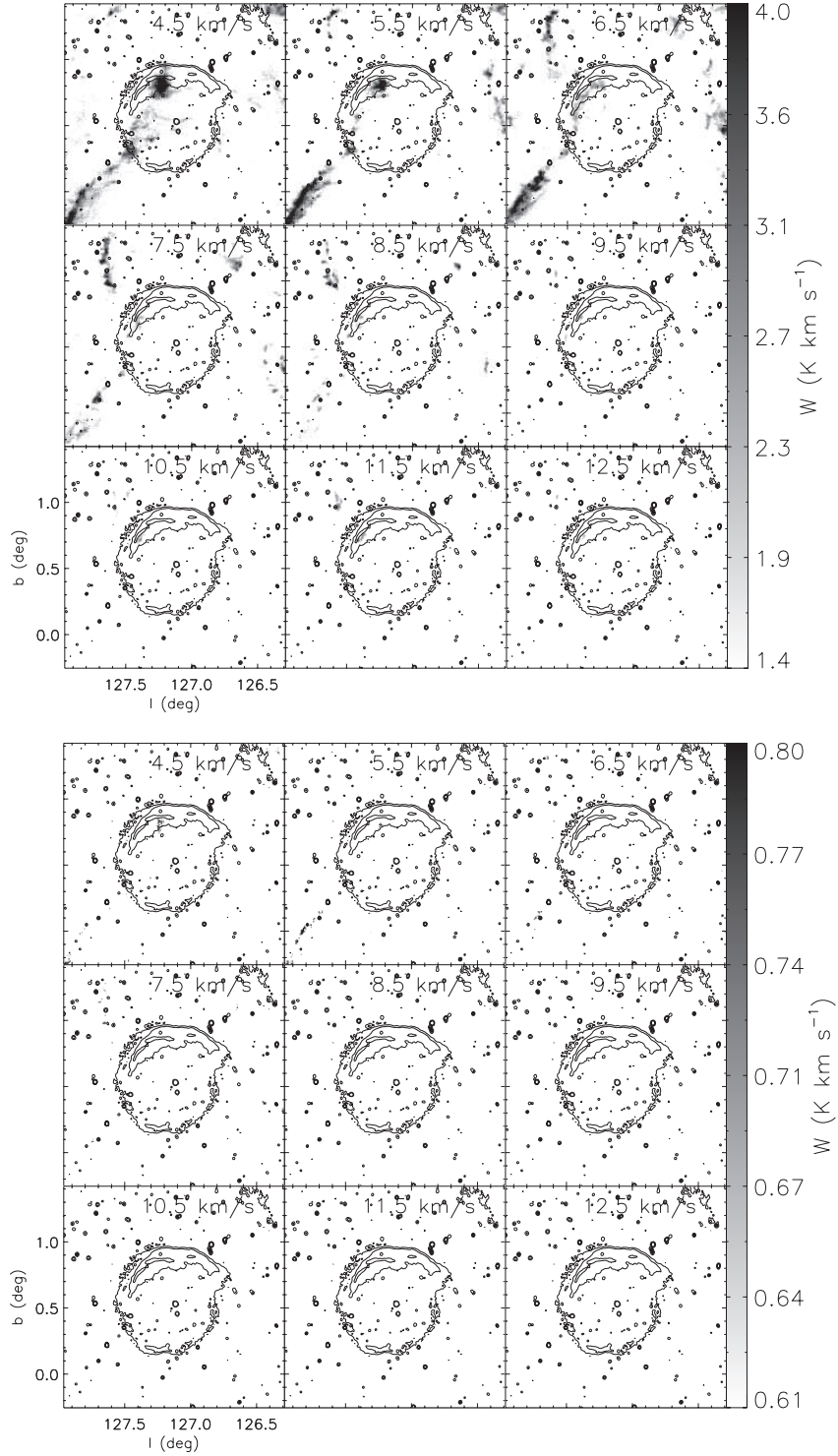,height=8.0in,angle=0, clip=}
\hfil\hfil}}
\caption{\twCO\ (J=1--0) (upper) and \thCO\ (J=1--0) (lower) emission maps  integrated over each 1~\km\ps, overlaid with contours of CGPS 1420~MHz radio continuum emission. The contour levels are 5.9, 6.9, and 8.8 K. Central velocities are indicated in each panel, and the minimum value of each map is $5\sigma$.}
\label{f:stamp1}
\end{figure*}

Figure~\ref{f:stamp1} shows the \twCO\ and \thCO\ intensity maps over the velocity range of 4--13~\km\ps\ with an interval of 1\km\ps.
No significant \CeiO\ emission was detected over any region.
There is a short \twCO\ filament in the northeastern region, with some corresponding \thCO\ emissions. Some diffuse \twCO\ emissions are also visible in the northwestern region, but there is no corresponding significant \thCO\ emission detected.
The prominent feature is a long filament that extends from the bottom-left corner to the central region. In comparison with the 1420~MHz continuum emission, we can divide the filament into two parts: a part coming out of the remnant region and a part in the remnant region.
The filament coming out of the remnant region is only present at velocities less than $\sim$10~\km\ps, and it is straight over all the velocities.
The filament in the remnant region is visible over all the velocities, and it is only straight at velocities less than $\sim$6~\km\ps.
At higher velocities, the filament is curved and coincident with the bright radio shell of the remnant. At the velocities over $\sim$10~\km\ps, we could only see the most curved part of the filament, which is at the position of the radio peak.
The \thCO\ emission of the long filament is mainly present at low velocities. There is no prominent \thCO\ emission in the curved part.

\begin{figure*}[ptbh!]
\centerline{{\hfil\hfil
\psfig{figure=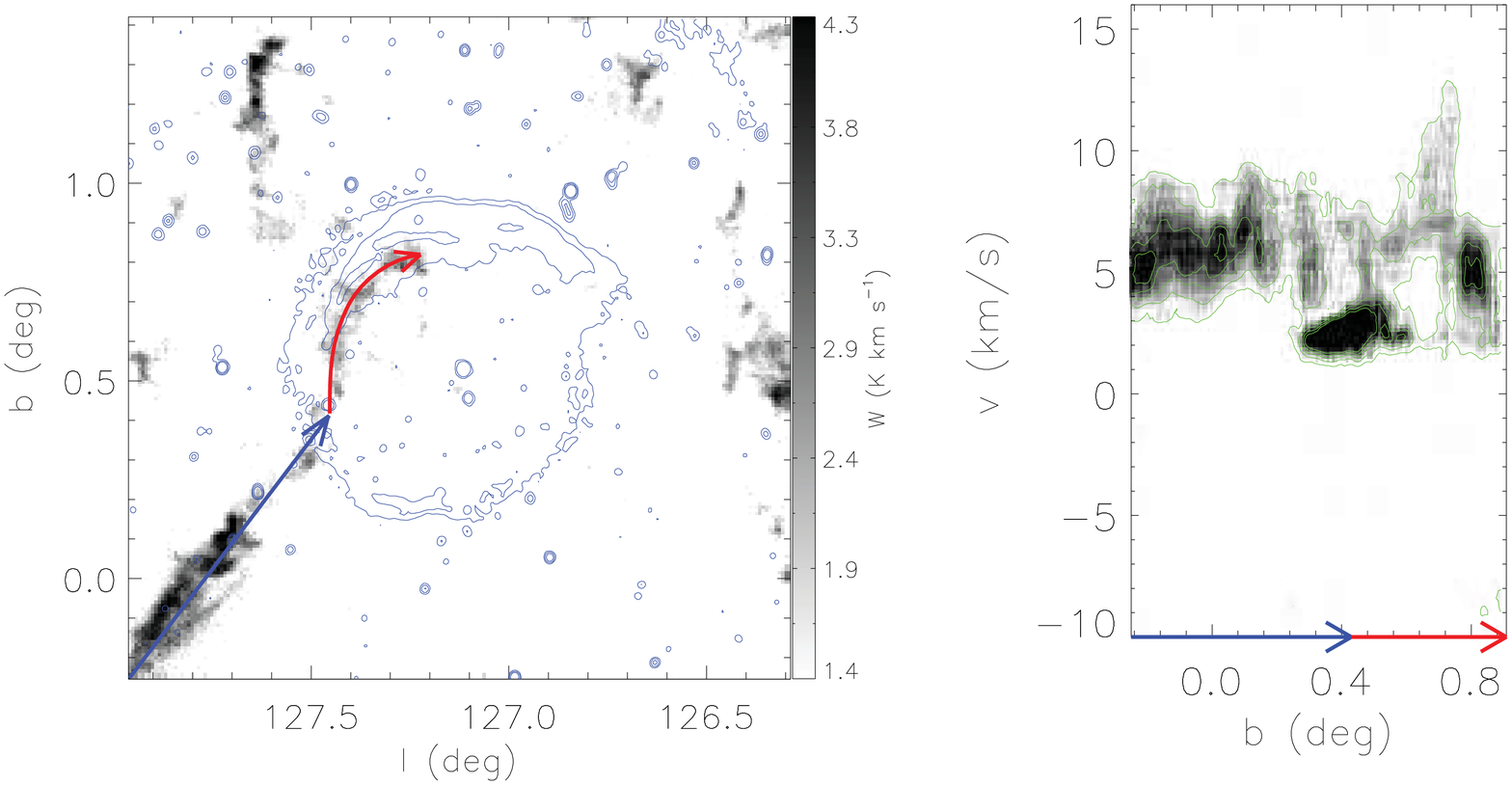,height=3.0in,angle=0, clip=}
\hfil\hfil}}
\caption{Left: integrated intensity map of \twCO\ (J=1--0) emission in the velocity range of 6--7~\km\ps, overlaid with the same 1420~MHz radio continuum contours as in Figure~\ref{f:stamp1}. The blue and red arrows indicate the straight and curved parts of the filament, which are out of the remnant and in the remnant, respectively.
Right: position-velocity map of \twCO\ (J=1--0) emission integrated over the strip region indicated by the arrows in the left panel, superposed with the intensity contours from $1\sigma$ (0.53~K) to $5\sigma$ in steps of $1\sigma$. The blue and red arrows correspond to that in the left panel.
}
\label{f:pvmap}
\end{figure*}

\begin{figure*}[ptbh!]
\centerline{{\hfil\hfil
\psfig{figure=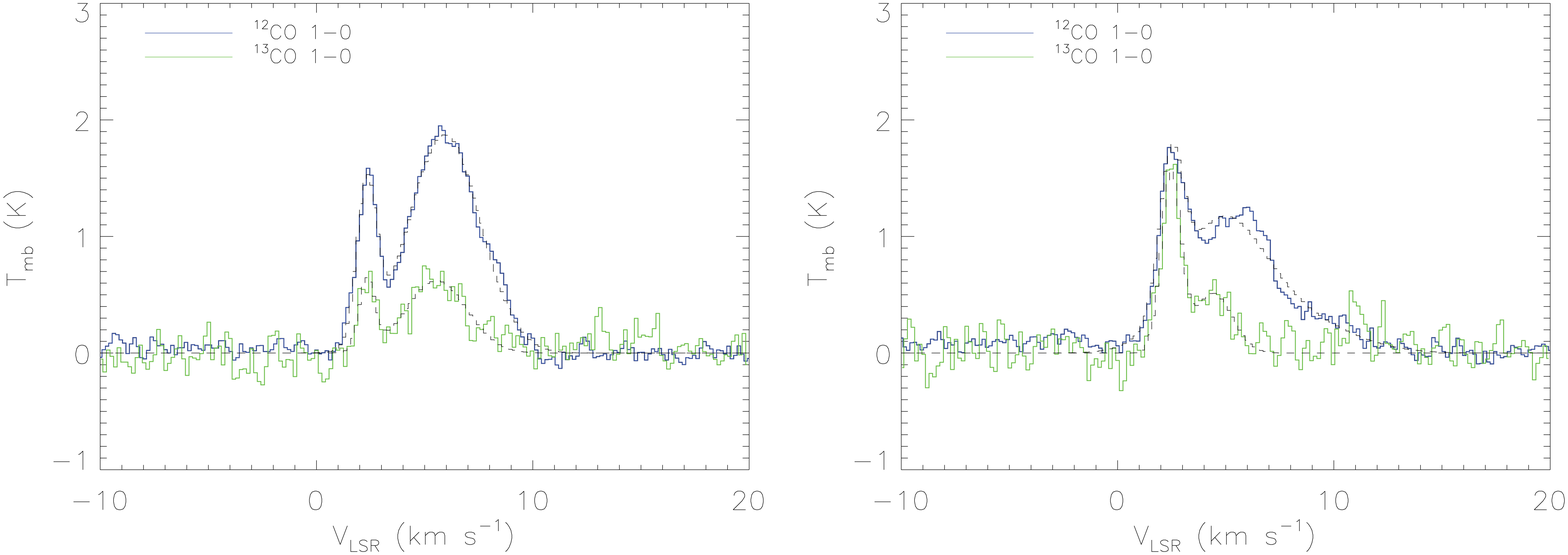,height=2.2in,angle=0, clip=}
\hfil\hfil}}
\caption{Average \twCO\ (J=1--0) and \thCO\ (J=1--0) spectra over the straight (left) and curved (right) parts of the filament, which are indicated by the blue and red arrows, respectively, in Figure~\ref{f:pvmap}.
The \thCO\ (J=1--0) spectra have been multiplied by 5 for a better visibility.
The spectra are fitted by two Gaussian components, except for the \twCO\ (J=1--0) spectrum of the curved part, which is fitted by three Gaussian components.
The fitting results are shown by the black dashed lines (see the fitted parameters in Table~\ref{tab:fitpara}).
}
\label{f:fitspec}
\end{figure*}

In the integrated intensity map of \twCO\ in the velocity range of 6--7~\km\ps, we could clearly see the long filament with the straight part coming out of the remnant region and the curved part in the remnant region, indicated by the blue and red arrows, respectively (Figure~\ref{f:pvmap}).
We found a compact broad redshifted CO wing that is only present in the curved part of the filament around ({\it l,b})=(127.4, 0.7), where the radio continuum shows a shell peak (see Figure~\ref{f:pvmap}).
This redshifted component is clearly visible in the \twCO\ (J=1--0) spectrum from the curved part of the filament (Figure~\ref{f:fitspec}).
These are strong evidences of the interaction between the SNR and the filament. Because of the compression and amplification of the magnetic field in such interaction, the radio peak can also be explained.
Such extensive CO wings have also been found in other SNRs, for example, SNR~W28 \citep{Arikawa+1999, Reach+2005} and SNR~IC~443 \citep{Dickman+1992, WangScoville1992, Seta+1998, Snell+2005}. However, in W28 and IC~443, the CO wings are very intense, and the velocity span could be over $\sim$50~\km\ps. It may indicate different amounts of shocked molecular gas (see Section~\ref{sec:phypar} for further discussion).

The position--velocity map along the whole filament shows some velocity distortions, with a velocity shift of $\sim$1~\km\ps. 
Similar velocity distortions have been found in other filamentary molecular clouds \citep[e.g.,][]{Bally+1987, KirK+2013, Zhang+2014}, which are probably common and may indicate a helical magnetic field in the filament \citep{Falgarone+2001}.
However, the situation is more complicated in the curved part of the filament, which contains both a broad redshifted wing around the most curved part and highly blueshifted bumps ($\sim$2~\km\ps) around the conjunction point between the straight part and the curved part. 
It suggests that such a velocity distortion could be enhanced by the impact of an SNR shock; in particular, the velocity shift will be enlarged by a shock propagating along the filament.

\begin{table*}\footnotesize
\begin{center}
\caption{Fitted Parameters of Thermal Molecular Lines for the Shocked Filament\label{tab:fitpara}}
\begin{tabular}{cccccccc}
\tableline\tableline
Region&Line&Peak \tmb &Center \vlsr&FWHM\\
&&(K)&(km \ps)&(\km \ps)\\
\tableline
Straight filament&\twCO\ (J=1--0)&1.87$\pm{0.02}$&5.93$\pm{0.03}$&3.90$\pm{0.06}$\\
&\thCO\ (J=1--0)&0.124$\pm{0.009}$&5.54$\pm{0.11}$&3.3$\pm{0.3}$\\
\tableline
Curved filament (rest)&\twCO\ (J=1--0)&1.05$\pm{0.11}$&4.8$\pm{0.7}$&4.60$\pm{0.61}$\\
&\thCO\ (J=1--0)&0.10$\pm{0.02}$&4.4$\pm{0.2}$&2.1$\pm{0.5}$\\
\tableline
Curved filament (redshifted)&\twCO\ (J=1--0)&0.29$\pm{0.02}$&8.11$\pm{0.21}$&6.0$\pm{0.5}$\\
&\thCO\ (J=1--0)&0$\pm{0.03}$\tablenotemark{a}&-&-\\
\tableline\tableline
\multicolumn{6}{c}{Derived parameters\tablenotemark{b}}\\
\tableline
Region&T$_{\rm ex}$&$\tau$(\thCO)&\ncol\tablenotemark{c}&M\tablenotemark{c}&M$_{\rm vir}$\tablenotemark{d}\\
&(K)&&(10$^{20}$ cm$^{-2}$)&($\Msun$)&($\Msun$)\\
\tableline
Straight filament&5.0&0.07&2.9 (14)&1.6$\E{2} {\rm d}^2_{1}$ (7.8$\E{2}$ d$^2_1$)\tablenotemark{e} &1.5$\E{4}$\du \tablenotemark{e} \\
Curved filament (rest)&4.0&0.1&2.0 (9.3)&70 ${\rm d}^2_{1}$ (3.2$\E{2}$ d$^2_1$)\tablenotemark{e} &1.3$\E{4}$\du \tablenotemark{e}\\
Curved filament (redshifted)\tablenotemark{f}&...&...&-- (3.3)&-- (1.1$\E{2}$ d$^2_1$)\tablenotemark{e} &2.2$\E{4}$\du \tablenotemark{e}\\
\tableline\tableline
\end{tabular}
\tablenotetext{a}{No \thCO\ emission visible, the error is the rms of the spectrum.}
\tablenotetext{b}{In the assumptions of local thermal equilibrium (LTE) and optically thick for \twCO\ emission.
Filling factor is assumed to be 1, since the width of the filament is much larger than the beam size.}
\tablenotetext{c}{Derived from \thCO\ column density by assuming the \thCO\ abundance of 1.4$\E{-6}$ \citep{Ripple+2013}, for comparison, we also show the values in the brackets, which are estimated by using the conversion factor ${\rm N(H}_2)/{\rm W(}^{12}{\rm CO)}\simeq1.8\times10^{20}~{\rm cm}^{-2}~{\rm K}^{-1}~{\rm km}^{-1}$~s \citep{Dame+2001}.}
\tablenotetext{d}{Calculated by $C\times{\rm k}_{2}\times{\rm L}\times\Delta v^{2}$, where k$_2$ is 105 \citep{MacLaren+1988}, ${\rm C}=0.65$ is the correction factor for filament-shape cloud, L is the length of the filament, and $\Delta v$ is the velocity width (FWHM).}
\tablenotetext{e}{\du=d/(1~kpc), where d is the distance of SNR \snr\ (see Section~\ref{sec:snr}).} 
\tablenotetext{f}{The assumption of optically thick for \twCO\ emission is not valid here.}
\end{center}
\end{table*}

We did not find any morphological correlations for the other velocity components other than the $\sim$5~\km\ps\ one in our observations.
In the velocity range of $-10$--20~\km\ps, there are two components, one is at $\sim$2~\km\ps\ and the other is at $\sim$5~\km\ps\ (see the left panel of Figure~\ref{f:fitspec}).
Both of these components are probably located in the local arm, but are from different subscale structures.
Note that the local standard of rest (LSR) velocity of the Perseus arm is around $-50$\km\ps\ in this direction \citep{Dame+2001}, and the distance is about 2~kpc \citep{Xu+2006, Hachisuka+2006}.
However, for a source at the LSR velocity of $-50$\km\ps, the kinematic distance is about 5~kpc by applying the rotation curve of \cite{BrandBlitz1993} together with the galactocentric distance of 8.0~kpc \citep{Reid1993} and the circular rotation speed of 220~\km\ps.
The reason for such discrepancy is the anomalous motion in this portion of the Perseus arm, which could be as large as 22\km\ps \citep{Xu+2006}.
Considering this peculiar motion, the velocity difference is still too large to place the velocity component at $\sim$5~\km\ps\ in the Perseus arm.
The LSR velocity in the local arm is around 0~\km\ps, and the maximum systematic velocity is $\sim$2~\km\ps\ \citep{BrandBlitz1993}, which is smaller than 5~\km\ps.
If we consider the contribution of an expanding shell near our solar system, which is known as the Gould Belt/Lindblad Ring, the $\sim$5~\km\ps\ systematic velocity could be understood; accordingly, the distance to this component would be about 300~pc \citep[][and references therein]{Bally2008}. 
Alternatively, for such a small velocity difference, it could also be accounted for by random cloud motion.
In this case, we could use half the distance to the Perseus arm, about 1~kpc, as a maximum.

\thCO\ (J=1--0) emissions are prominent for these two velocity components, with their line centers consistent with that of \twCO\ (J=1--0) emission, while the \twCO\ emission from the curved part of the filament is broadened in the redwing (see the right panel of Figure~\ref{f:fitspec}), and there is no corresponding \thCO\ emission detected for the broadened redwing.
In fact, the peak of the \twCO\ emission is also shifted a little relative to the peak of the \thCO\ emission in addition to the red-wing broadening.
Considering that the absence of corresponding \thCO\ emission contributes most of the uncertainty in our spectral analysis, we used only one component to fit the \twCO\ redshifted component.
The upper limit of the peak ${\rm T}_{\rm mb}$ for the \thCO\ emission of the redshifted component is 0.03~K and the upper limit of full width at half maximum (FWHM) is 6.0~\km\ps, which are the rms of the spectrum and the FWHM of the corresponding \twCO\ component, respectively.
Except for the \twCO\ (J=1--0) spectrum from the curved part of the filament fitted by the three Gaussian components, we used two Gaussian components to fit all other spectra (see the fitting results in Figure~\ref{f:fitspec}). The fitted and derived parameters are listed in Table~\ref{tab:fitpara}.

We used two methods to estimate the H$_2$ column density and the molecular mass.
In the first method, we assumed local thermodynamical equilibrium (LTE) for the gas and optically thick conditions for the \twCO\ (J=1--0) line and used the \thCO\ abundance of 1.4$\E{-6}$ \citep{Ripple+2013}.
For comparison, we also estimated the H$_2$ column density by applying the conversion factor ${\rm N(H_2)/W(^{12}CO)}\simeq1.8\times10^{20}~{\rm cm}^{-2}~{\rm K}^{-1}~{\rm km}^{-1}$~s \citep{Dame+2001}.
The differences between the H$_2$ column densities derived by the two methods are partly caused by the small filling factors of the \thCO\ emission, since there are less \thCO\ emissions than \twCO\ emissions in the filament.
The effect is not significant, however, because the H$_2$ column density will not change much in the case of a small filling factor. Applying the possible filling factor of \thCO\ emission of 0.15, which is the ratio of the number of points with bright \twCO\ emission and the number of points with bright \thCO\ emission, the H$_2$ column density of the straight part of the filament becomes 3.7$\E{20}$~cm$^{-2}$.
The derived molecular gas masses for the filament are all about two orders lower than the virial masses, indicating that the filamentary molecular cloud is not confined by gravity \citep[maybe by magnetic field;][]{Contreras+2013}.

\section{Discussion}\label{sec:discuss}
\subsection{Physical Conditions of the Molecular Gas}\label{sec:phypar}

\begin{figure}[ptbh!]
\centerline{{\hfil\hfil
\psfig{figure=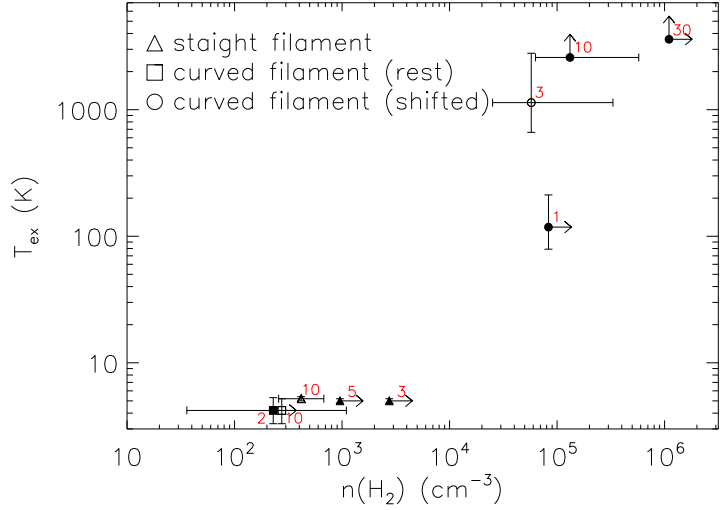,height=3in,angle=90, clip=}
\hfil\hfil}}
\caption{Physical parameters derived from the fitted \twCO\ (J=1--0) and \thCO\ (J=1--0) line profiles, according to the RADEX radiative transfer modeling.
Triangles denote the results for the $\sim$5~\km\ps\ rest velocity component of the straight part of the filament, squares for the $\sim$5~\km\ps\ rest velocity component of the curved filament, and circles for the shifted broad velocity component of the curved filament.
The lower limits are shown by filled symbols.
All symbols are labeled by the H$_2$ column densities in units of $10^{20}$~cm$^{-2}$.
In the calculation, we assume the isotopic abundance of \twCO/\thCO=70 \citep{Milam+2005}.
}
\label{f:fitpara}
\end{figure}

According to the fitted parameters of the molecular filament (see Table~\ref{tab:fitpara}), we ran the RADEX radiative transfer code \citep{vanderTak+2007} with the LAMDA molecular data files \citep{Schoeier+2005} to simultaneously model the \twCO~(J=1--0) and \thCO~(J=1--0) line emissions.
The two molecular lines are sensitive to different physical parameters, hence the joint modeling could help constrain the physical state of the molecular gas.
In the modeling, we chose the slab option in RADEX to mimic the geometries of the shocked and filament-shaped molecular clouds, and assumed a \thCO\ abundance of 1.4$\E{-6}$ \citep{Ripple+2013} and an isotopic abundance of \twCO/\thCO=70 \citep{Milam+2005}.
In running RADEX, we set the collisional partner to be H$_2$ and set the temperature of the background radiation to be the cosmic microwave background (CMB) temperature, that is, 2.73~K.
We explored the physical parameter space over excitation temperature and density to find the best-fit model and then calculated 90\% confidence intervals on the best-fit parameters by applying the $\chi$-squared statistics. 
We estimated the best-fit results using different plausible column densities; however, we could only get the lower limits for some of them.
Figure~\ref{f:fitpara} shows the best-fit parameters and their confidence intervals.

We could distinguish two kinds of molecular gases in Figure~\ref{f:fitpara}: one is cold and dense from the straight molecular filament and the rest component toward the curved molecular filament, which is consistent with quiet molecular clouds;
the other is hot and about two orders of magnitude denser from the redshifted component toward the curved filament, which is consistent with a shocked molecular cloud.

The fitting results of quiet molecular gases for the column densities lower than $10^{21}$~cm$^{-2}$ only give lower limits of number densities. Then, the lengths in the line of sight (LoS) could be derived as N(H$_2$)/n$<0.2$~pc, which is too small compared to the width of the filament ($\sim$2\du~pc).
Further, the corresponding lengths in the LoS are $\sim$1~pc, which is consistent with the width of the straight filament.
These column densities are consistent with those derived 
from the assumptions of LTE and \twCO\ (J=1--0) optically thick and from the using the conversion factor (see Table~\ref{tab:fitpara}).

The best-fit result for the redshifted component gives a density of $\sim$6$\E{4}$~cm$^{-3}$ and a temperature of $\sim$1000~K with the column density of $3\E{20}$~cm$^{-2}$.
Note that, for the redshifted component, the optical depth of the \twCO\ (J=1--0) line emission is not large, because of the large velocity gradient (broadened line) and non-detection of corresponding \thCO\ (J=1--0) line emission, so the temperature could not be constrained well here.
Besides, the molecular data are not enough to handle temperatures higher than 500~K in running RADEX.
In our fitting result, the temperature ranges from a hundred to over a thousand Kelvin. 
Considering the non-detection of OH~(1720 MHz) line emission \citep{Frail+1996}, the temperature is probably higher than 125~K \citep{WardleYusef-Zadeh2002}. However, the missing OH~(1720 MHz) line emission may also indicate a low column density like that in a J-type shock.
In the case of weakly magnetized shock, i.e., a J-type shock, we can derive the post-shock temperature as $\zeta \bar\mu m_{\rm H} v_{\rm cloud}^2 / k_{\rm B}$, where $\zeta=3/16$ for a large Mach number, $\bar\mu=2.33$ is the average atomic weight \citep{Lesaffre+2013}, $m_{\rm H}$ is the mass of the hydrogen atom, $v_{\rm cloud}$ is the velocity of the shock in the filament, and $k_{\rm B}$ is Boltzmann constant.
The molecular shock velocity is $v_{\rm cloud}=4 v_{\rm shift}/3\sim5$~km s$^{-1}$, where $v_{\rm shift}$ is the velocity of post-shock molecular gas. Hence, the calculated post-shock temperature is $\sim$1300~K, which is consistent with the best-fit result.
If the shock is highly magnetized, i.e., a C-type shock, the post-shock temperature would be much smaller than that in a J-type shock, because the heating is spread out over a much larger region, especially for a slow shock \citep{Lesaffre+2013}.
Therefore, the molecular gas is probably heated and compressed by a non-dissociative J-type shock, but the existence of a C-type shock cannot be excluded.

Adopting the best-fit column density and the number density of the redshifted component, we have the length in the LoS of $l\sim1.6\E{-3}$~pc,
which is about three orders of magnitudes less than the width of the straight filament.
Since the thickness of shocked molecular gas (represented by $s$) is less than the length in the LoS ($s\lesssim l$),
which is very small,
the shocked molecular gas is probably in a very thin sheet.
We consider that all shocked molecular gas is compressed into this thin sheet, which is reasonable for the slow shock. Therefore, the thickness of the corresponding unshocked gas is $s_0=s\times n_{\rm shock}/n_{\rm cloud}\lesssim l\times n_{\rm shock}/n_{\rm cloud}\sim0.32$~pc, where $n_{\rm shock}$ is the density of shocked molecular gas, and $n_{\rm cloud}$ is the density of quiet molecular gas,
which is approximately one-sixth of the width of the straight filament.
The time for the shock to pass through such a distance is $t_{\rm cloud}=s_0/v_{\rm cloud}\lesssim 6\E{4}$~yr.
Applying the column density of 3$\E{20}$~cm$^{-2}$, we obtain the mass of the shocked molecular gas to be $100$\du$^{2}~M_\odot$ and the kinematic energy of $\sim$1.6$\E{46}$\du$^2$~erg which provides a lower limit for the SNR energy.

The amount of shocked molecular gas is remarkably less than that in W28 and IC~443. 
In W28, there are $\sim$2$\E{3}~\Msun$ shocked molecular gases, with the total kinetic energy of $\sim$3$\E{48}$~erg \citep{Arikawa+1999}.
In IC~443, there are 500--2000~$\Msun$ shocked molecular gases, with a possible total kinetic energy of $\sim$3--13$\E{48}$~erg \citep{Dickman+1992}.
Both W28 and IC~443 are interacting with large MCs ($\sim$4$\E{3}~\Msun$ for W28, \citeauthor{Arikawa+1999}~\citeyear{Arikawa+1999}; $\sim$1$\E{4}~\Msun$ for IC~443, \citeauthor{Cornett+1977}~\citeyear{Cornett+1977}), and their shock fronts are blocked by the surrounding MCs.
The case is different in \snr, which is interacting with a part of a small filamentary MC that could be engulfed in the remnant.
The overall densities and temperatures of quiet molecular gases associated with W28 ($\sim$10$^{3}~\cm^{-3}$ and $\sim$20~K, respectively; \citeauthor{Arikawa+1999}~\citeyear{Arikawa+1999}) and \snr\ are similar. However, the detailed structures of the associated MCs may be very different.
The properties of their shocked molecular gases are found to be different, taking into account that W28 is associated with 1720~MHz OH masers \citep{Claussen+1997} and \snr\ is not.
The origin of this difference is not clear. It needs more observations for further investigation.

\subsection{Nature of SNR \snr}\label{sec:snr}

Both morphological correspondence and dynamical evidence indicate an association between SNR~\snr\ and the molecular filament.
We also examined the H{\sc i} 21~cm line emission data from CGPS, and found no absorption feature toward the remnant, which is consistent with the location of the molecular filament at the far side of the remnant (red-wing broadening).
This association facilitates us to determine the kinematic distance to the remnant.
As shown above, the molecular filament is in the local arm, so is the remnant.
If the molecular filament is in the Gould Belt/Lindblad Ring, the distance of the SNR is ${\rm d}\sim300$~pc, and the radius is $r_{\rm s}\sim2.3$~pc.
Moreover, the detection of optical emission probably enhances our results of shock--cloud interaction and the close distance; however, the overlapping of the open cluster NGC~559 and the nearby northeastern H{\sc ii} region prohibit a detailed morphological study \citep{Xilouris+1993}.
The distance to the open cluster NGC~559 is $\sim$1.3~kpc \citep{RoseHintz2007}, and there is an H{\sc i} bubble at $-14$~\km\ps\ surrounding it, of which the kinematic distance is consistent with that of NGC~559 \citep{LeahyTian2006}.
Based on our observations, there is also a CO cloud at $\sim$$-$13~\km\ps, but with a smaller extension than H{\sc i} cloud, which is probably the molecular counterpart of H{\sc i} gas. 
The bubble structure shown in H{\sc i} emission cannot be distinguished in our CO observations, which might be due to the limited sensitivity.
The distance to the NGC~559 and H{\sc i} bubble system is about half the distance to the Perseus arm. It shows that the local arm in this direction could be quite extended.
We cannot exclude the association of the SNR with NGC~559 as well as with the H{\sc i} bubble, although we have not found any H{\sc i} absorption feature toward the remnant.
If they are associated, the distance of the SNR is ${\rm d}\sim1.3$~kpc, and the radius is $r_{\rm s}\sim10$~pc.

There are radio emissions beyond the filament in the northeastern region, which may be caused by a projection effect.
In an extreme case, if the filament is located at the boundary of the remnant, 
the velocities between the forward shock and the molecular shock should be similar, since the two shocks have passed through about the same distance in the same time ($t_{\rm cloud}\lesssim6\E{4}$~yr as shown above).
Consequently, the density contrast is $\chi=n_{\rm cloud}/n_{0}\sim \beta v_{\rm s}^2/v_{\rm cloud}^2\sim1$, where $n_0$ is the density of the circumstellar or interstellar medium (CSM/ISM), $v_{\rm s}$ is the forward shock velocity, and the constant $\beta$ is adopted as unity \citep{McKeeCowie1975,Orlando+2005}.
If the density of the CSM/ISM is actually as large as $n_{\rm cloud}$ ($\sim$300~cm$^{-3}$), the accumulated gas swept up by the remnant shock will form a massive shell of about 2.8$\E{4}~\Msun$, which is much larger than the mass of the filament. We have not found such a shell in our observations.
On the other hand, if the radial direction is approximately perpendicular to the LoS,
the forward shock should have been surrounding the filament after the encounter, and the forward shock will propagate about 10\du\ times farther away than the molecular shock after the encounter. Therefore, its velocity is also about 10\du\ times larger than the molecular shock, which is $\sim$50\du~\km\ps.
In this case, the density contrast is $\chi\sim50$\du, and the density of the CSM/ISM is $n_{0}=n_{\rm cloud}/\chi\sim6$\du$^{-1}$~cm$^{-3}$, which is more reasonable.
Therefore, the velocity of the forward shock is probably $v_s\sim50$\du~\km\ps, and the corresponding post-shock temperature is $\sim$10$^{5}$~K.
The age of the remnant is greater than $t_{\rm cloud}$, which is $\lesssim8\E{4}$~yr. It is not well constrained yet, but this still indicates that the remnant is relatively old compared to a young SNR, e.g., SNR~W49B \citep{Zhou+2011}.

The low post-shock temperature indicates that the remnant is in the radiative phase, which is also confirmed by the non-detection of \xray\ emission in \ROSAT\ All-Sky survey data.
The age of the remnant in the radiative phase can be estimated as ${\rm t}=2 r_{\rm s}/(7 v_{\rm s})\sim4.3\E{4}$~yr \citep{McKeeOstriker1977}, and the result is consistent with that shown above.
Based on the numerical solutions and analytical fitting for an SNR in the pressure-driven snowplough phase by \cite{Cioffi+1988}, the explosion energy could be obtained as:
\begin{equation}
 E=6.8\E{43}{\rm n_0}^{1.16}(\frac{v_s}{1~{\rm km}~{\rm s}^{-1}})^{1.35}(\frac{r_{\rm s}}{1~{\rm pc}})^{3.16}\zeta_m^{0.161}~{\rm erg} \sim 6.8\E{49}{\rm d}_1^{3.35}~{\rm erg},
\end{equation}
where $\zeta_m={\rm Z/Z}_\odot$ is the metallicity parameter, which is set to 1.
The explosion energy, 1.6$\E{50}$~erg, derived by applying the distance of 1.3~kpc, is lower than the nominal value ($10^{51}$~erg), and it is 1.2$\E{48}$~erg if the distance is 300~pc, which is probably too small.
It may support the large distance of the remnant.
Evidences of such sub-energetic supernova explosion have been found in other SNRs, e.g., SNR~DA530 \citep{Jiang+2007}, SNR~3C~58 \citep{RudieFesen2007}, and SNR~G310.6-1.6 \citep{Renaud+2010}, which could originate from the core-collapse supernova explosion of relatively low mass ($\sim$8--10~M$_{\odot}$) stars \citep{Nomoto1987}.

\subsection{Possible Associated Star Formation}

\begin{figure}[ptbh!]
\centerline{{\hfil\hfil
\psfig{figure=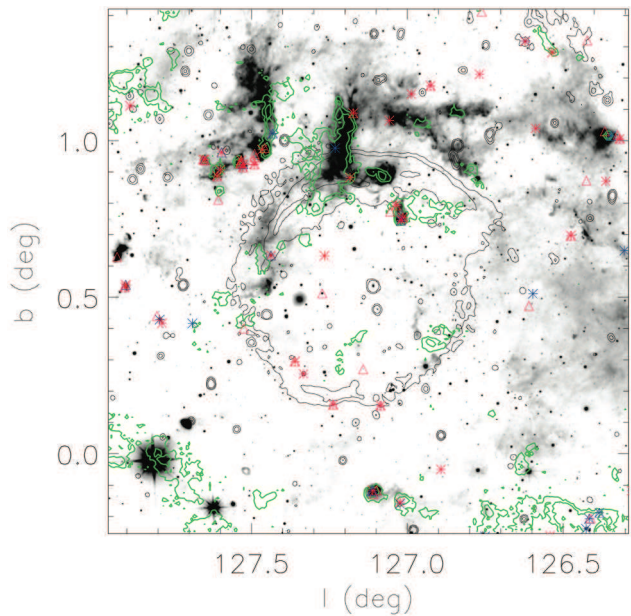,height=3.2in,angle=90, clip=}
\hfil\hfil}}
\caption{Image of the 12~$\mu$m emission observed by {\it WISE}, overlaid with
the contours of the 1420~MHz radio continuum emission (black; the same as in Figure~\ref{f:stamp1}) and the \twCO\ (J=1--0) emission in the velocity range of $-60$ to $-35$~\km\ps (green).
The positions of young stellar objects (YSOs) in the SNR region are labeled, with type Class~I in blue and type Class~II in red.
The stars denote the YSOs selected from the first three bands of {\it WISE} all-sky survey data, and the triangles those from the first two bands of {\it WISE} data and the H and Ks bands of 2MASS all-sky survey data.
The \twCO\ emission contours are $5\sigma$ and $10\sigma$, where the rms is 6.65~K~\km\ps.
}
\label{f:yso}
\end{figure}

There is corresponding mid-IR emission from the curved filament, but no mid-IR emission is seen from the straight filament (see Figure~\ref{f:yso}).
These mid-IR emissions probably originate from the hot dust that is heated by the remnant shock.
There is also mid-IR emission in the northeastern region outside the remnant, which may originate from the hot dust related to the nearby H{\sc ii} region G127.5+1.1 \citep{FichTerebey1996}.

Using the IR data we selected the young stellar objects (YSOs)
to investigate their spatial distribution.
In our examination, we only focus on the disk-bearing young stars, since their IR colors are distinctly different from those of diskless objects. We cannot distinguish diskless young stars from unrelated field objects based on IR colors alone.
Following the criteria described in \cite{Koenig+2012}, we selected candidate disk-bearing young stars, where the contaminations from extragalactic sources (star-forming galaxy and active galactic nucleus), shock emission blobs, and polycyclic aromatic hydrocarbon (PAH) emission objects were removed based on their locations in the [3.4]$-$[4.6] versus [4.6]$-$[12] color--color diagram and their {\it WISE} photometry.
For the sources that are not detected in the {\it WISE} [12] band,
we constructed the  $K_{\rm s}$$-$[3.4] versus [3.4]$-$[4.6] color--color diagram based on their dereddened photometry in the {\it WISE} [3.4] and [4.6] bands, in combination with the dereddened 2MASS $K_{\rm s}$ photometry.
To deredden the photometry, we estimated the extinction by their locations in the $J$$-$$H$ versus $H$$-$$K_{\rm s}$ color--color diagram as described in \cite{Fang+2013}.

\begin{figure}[ptbh!]
\centerline{{\hfil\hfil
\psfig{figure=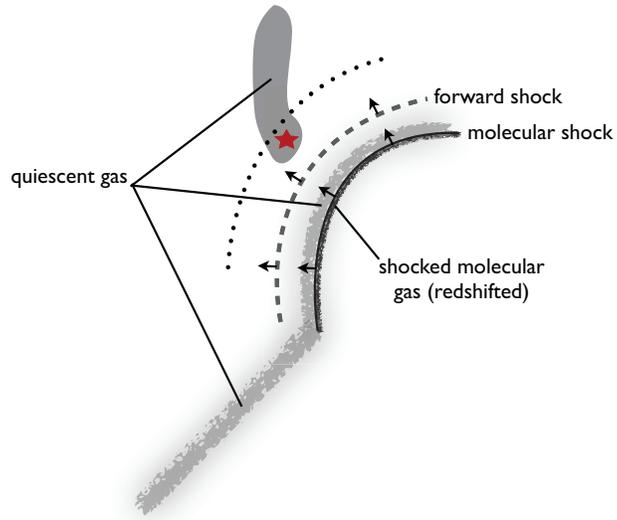,height=3.2in,angle=0, clip=}
\hfil\hfil}}
\caption{
Schematic view of SNR~\snr\ based on our \twCO\ (J=1--0) and \thCO\ (J=1--0) observations.
The red star denotes the position of the assembled YSOs (see Figure~\ref{f:yso}).
The dotted line denotes the possible border of the progenitor's stellar wind.
}
\label{f:schematic}
\end{figure}

We found that the majority of aggregated YSOs are coincident with the MCs at $\sim$$-50$~\km\ps\ that are located in the Perseus spiral arm at a distance of $\sim$2~kpc \citep{Xu+2006}, except that a cluster of YSOs at ($l\sim127^\circ.5$, $b\sim0^\circ.93$) shows no morphological correlation with the velocity component at $\sim$$-50$~\km\ps, and no association to the mid-IR emission either.
Instead, this cluster of YSOs is located at the top end of the northeastern molecular filament at $\sim$5~\km\ps\ (see Figure~\ref{f:stamp1}).
There is no morphological correlation found between the cluster of YSOs and other velocity components.
It suggests a star-forming activity at the top end of the northeastern molecular filament close to the SNR.
Since the shock of the remnant does not reach yet here, the star-forming activity cannot be related to the remnant shock.
Nevertheless, the stellar wind of the progenitor of the remnant could trigger star formation here (see Figure~\ref{f:schematic}).
The lack of mid-IR emission as well as \thCO\ emission around the YSOs may indicate that the diffuse tenuous gas was evacuated by the stellar wind that triggered the collapse of the dense gas in the meantime.

\section{Conclusions}\label{sec:conclusion}

We have performed millimeter observations in CO lines toward SNR~\snr. We found a molecular filament at 4--13~\km\ps\ that consists of two distinct parts: a straight part coming out of the remnant region and a curved part in the remnant region.
The deviation from virial equilibrium suggests that the filament is not confined by gravity (maybe by magnetic field).
The curved part is well consistent with the bright SNR shell detected in the 1420~MHz radio continuum and mid-IR observations.
Further, redshifted line wing broadening is also found only in this curved part of the molecular filament, which indicates a physical interaction.
These provide strong evidences, for the first time, to confirm the association between an SNR and a pre-existing long molecular filament.
Multiband observations in the northeastern remnant shell could be explained by the interactions between the remnant shock and the dense molecular filament.
With the association between the SNR~\snr\ and the molecular filament at $\sim$5~\km\ps, we could place the SNR in the Gould Belt/Lindblad Ring and at a kinematic distance of about 300~pc. 
Alternatively, there is also another possibility that the molecular filament has a peculiar motion, and the remnant may be associated with the open cluster NGC~559 and located at a farther distance of about 1.3~kpc. It is supported by the estimation of explosion energy.

Radiative transfer modeling of the physical conditions of the molecular gas in the filament in combination with the assumed columnar-shaped filament yields a column density of $\sim$$10^{21}$~cm$^{-2}$, a density of $\sim$300~cm$^{-3}$, and a temperature of $\sim$5~K for the quiet component. For the shocked component in the curved part of the filament, the physical parameters were found to have a column density of $\sim$3$\E{20}$~cm$^{-2}$, a density of $\sim$6$\E{4}$~cm$^{-3}$, and a temperature of $\sim$1000~K. The shocked component probably originates from the passage of a non-dissociative J-type shock.
The interaction details could be well explained if the curved part of the filament is fully engulfed by the remnant forward shock, and it indicates a dense CSM/ISM environment surrounding the SNR ($\sim$6\du~cm$^{-3}$) and a low explosion energy ($\sim$6.8$\E{49}$\du$^{3.35}$~erg).
It also suggests that the SNR is in radiative phase, and the age is estimated to be $\sim$4.3$\E{4}$~yr.
We also found a spatial correlation between the aggregated YSOs and the adjacent molecular filament close to the SNR, which could be triggered by the progenitor's stellar wind.

\acknowledgments
We are grateful to all the members of the Milky Way Scroll Painting-CO line survey group, especially the staff of Qinghai Radio Observing Station at Delingha for the support during the observation.
We thank the anonymous referee for providing very helpful comments that improved the paper and its conclusions.
This work is supported by NSFC grants 11233007 and 11233001, and by the Jiangsu Provincial Natural Science Foundation, Grant No. BK20141044.
M.F. acknowledges support by the NSFC through grants 11203081, and Y.S. acknowledges support by the NSFC through grants 11103082.
This work is also supported by the Strategic Priority Research Program of the Chinese Academy of Sciences, Grant No. XDB09000000.
We acknowledge the use of VGPS data; the National Radio Astronomy Observatory is a facility of the National Science Foundation operated under cooperative agreement by Associated Universities, Inc.
This publication makes use of data products from the Two Micron All Sky Survey, which is a joint project of the University of Massachusetts and the Infrared Processing and Analysis Center/California Institute of Technology, funded by the National Aeronautics and Space Administration and the National Science Foundation.
This publication makes use of data products from the {\it Wide-field Infrared Survey Explorer}, which is a joint project of the University of California, Los Angeles, and the Jet Propulsion Laboratory/California Institute of Technology, funded by the National Aeronautics and Space Administration.

\bibliographystyle{apj}
\bibliography{/Users/zx/back/Literature/documents/squeezer/template/tex/mine/ref}

\end{CJK*}
\end{document}